\documentclass[conference]{IEEEtran}
\IEEEoverridecommandlockouts
\usepackage{cite}
\usepackage[hyphens]{url}
\usepackage{amsmath,amssymb,amsfonts}
\usepackage{algorithmic}
\usepackage{graphicx}
\usepackage{textcomp}
\usepackage{multirow}
\usepackage{xcolor}
\def\BibTeX{{\rm B\kern-.05em{\sc i\kern-.025em b}\kern-.08em
    T\kern-.1667em\lower.7ex\hbox{E}\kern-.125emX}}

\newcommand\blfootnote[1]{%
  \begingroup
  \renewcommand\thefootnote{}\footnote{#1}%
  \addtocounter{footnote}{-1}%
  \endgroup
}

\begin{document}

\title{Progressive Searching for Retrieval in RAG}


\author{\IEEEauthorblockN{Taehee Jeong, Xingzhe Zhao, Peizu Li, Markus Valvur, Weihua Zhao}
\IEEEauthorblockA{\textit{San Jose State University} \\
taehee.jeong, xingzhe.zhao, peizu.li, markus.valvur, weihua.zhao@sjsu.edu}

{\thanks{This work was supported in part by a Mobilint Grant awarded to San Jose State University. (Corresponding author: Taehee Jeong)}
}
}

\maketitle

\begin{abstract}
Retrieval-Augmented Generation (RAG) is a promising technique for mitigating two key limitations of large language models (LLMs): outdated information and hallucinations. RAG system stores documents as embedding vectors in a database. Given a query, search is executed to find the most related documents. Then, the topmost matching documents are inserted into LLMs’ prompt to generate a response. Efficient and accurate searching is critical for RAG to get relevant information. We propose a cost-effective searching algorithm for retrieval process. Our progressive searching algorithm incrementally refines the candidate set through a hierarchy of searches, starting from low-dimensional embeddings and progressing into a higher, target-dimensionality. This multi-stage approach reduces retrieval time while preserving the desired accuracy. Our findings highlight the trade-offs between dimensionality, speed, and accuracy and demonstrate that progressive search enables scalable, high-performance retrieval for large databases in RAG systems. 
Our findings demonstrate that progressive search in RAG systems achieves a balance between dimensionality, speed, and accuracy, enabling scalable and high-performance retrieval even for large databases.
Our code is available at https://github.com/taeheej/Progressive-searching-for-Retrieval-in-RAG.

\end{abstract}


\section{Introduction}
\renewcommand{\footnoterule}{%
  \kern -3pt
  \hrule width 3in height 1pt
  \kern 2pt
}
\blfootnote{24th International Conference on Machine Learning and Applications (ICMLA), 
IEEE Copyright 2025}

In recent years, the growth of Large Language Models (LLMs) has significantly advanced natural language processing (NLP), particularly in the domain of information retrieval and question answering. Retrieval-Augmented Generation (RAG) systems have emerged as a powerful methodology that enhances the performance of LLMs by leveraging external knowledge bases \cite{b1}.

The effectiveness of a RAG system fully depends on the quality of the document retrieval process, which is inherently influenced by the embedding model used to represent documents and queries. High-dimensional vector embeddings, which capture the semantic meaning, are central to the retrieval process; however, their effectiveness is dependent on the embedding model used and the level of dimensionality they encompass \cite{b2} \cite{b3}.

The existing literature provides limited insight into how embedding dimensionality impacts retrieval performance in large-scale systems. Furthermore, while conventional Nearest Neighbors (KNN) \cite{b4}-based retrieval can achieve high accuracy, it often suffers from performance issues with regard to speed when operating on high-dimensional vectors across large datasets. Essentially, there are not many guidelines with respect to how to balance dimensionality, retrieval accuracy, and system efficiency. This poses an issue in our modern era as models are becoming increasingly complex and require further resources in order to perform with the efficiency that is desired.

\begin{figure}[ht]
\centerline{\includegraphics[width=0.95\linewidth]{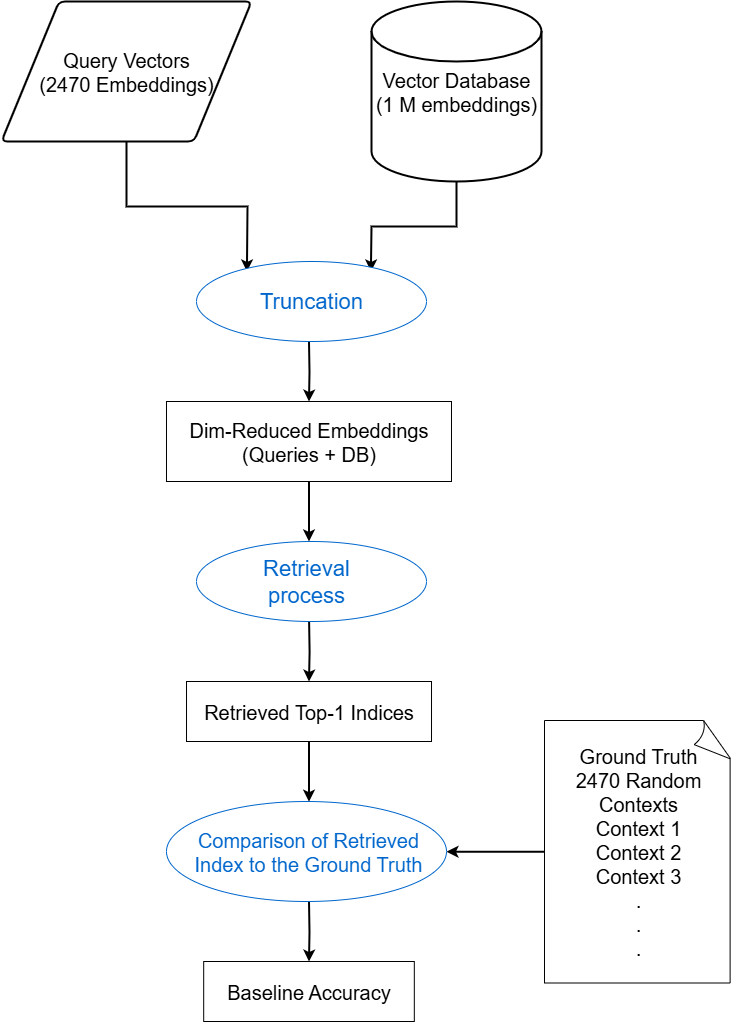}}
\caption{Truncated Retrieval workflow}
\label{fig:retrieval}
\end{figure}

We address these gaps by evaluating retrieval accuracy for various dimensionalities using embedding from OpenAI and Alibaba-NLP across a one million document corpus. In addition to this, 
we propose a progressive searching algorithm. This method performs retrieval in multiple stages. Beginning with a low-dimensional search to narrow down candidate documents, followed by one or more higher-dimensional search passes on the smaller subsets, ultimately retrieving final results for the desired dimensionality. By tuning parameters such as step size, starting low dimensionality, and desired dimensionality, we demonstrate that this approach can reduce query times while preserving top-1 retrieval accuracy.

Our work has provided a comprehensive understanding of how embedding dimensionality and retrieval strategies interact. The insights gained inform best practices for building scalable and efficient RAG pipelines, especially in the context of large-scale document retrieval scenarios.

\section{Related Work}
RAG has shown significant power in enhancing large language models by integrating different embedding models. Embedding models are the fundamental component for RAG because they decide how information is put into the system by translating texts to numerical vectors. 
For the embedding models, we used OpenAI' text-embedding-3-large \cite{b5} and Alibaba-NLP's gte-Qwen2-7B-instruct \cite{b6}. The output dimensions are 3072 and 3584 for OpenAI and Alibaba-NLP embedding models respectively.

In the context of RAG, embedding models usually generate vector representations in extremely high dimensions, making computation rather long and expensive. These dimensions can go beyond hundreds and thousands. Principal Component Analysis (PCA) \cite{b7} focus on those dimensions with actual information instead of noise while preserving the greatest variance. It identifies orthogonal directions (principal components) along which the variance in the data is maximized, allowing high-dimensional data to be projected into a lower-dimensional subspace while preserving as much variance as possible.  
Though helpful, PCA has its own flaws. 
Since PCA projects data onto a linear subspace, it cannot capture the non-linear structure that often exists in real-world datasets. 
In addition, PCA is very sensitive to outliers. A small outlier might be able to twist the principal components in the wrong direction. In the context of RAG, PCA is rather incompetent in capturing non-linear correlations in the high dimensionality generated by complex embedding models.

KNN is a classical algorithm for pattern recognition, classification, regression, and similarity-based retrieval tasks. KNN algorithm identifies the k closest data points to a given query based on a distance metric, such as Euclidean or cosine similarity, and derives the output by aggregating the labels or properties of the neighbors. Due to its simplicity, interpretability, and effectiveness, KNN is a popular baseline in information retrieval. In this work, KNN served as a baseline searching algorithm with truncation for comparison purposes. The progressive searching method discussed in this work was inspired by the KNN method.

Distance metrics, such as Euclidean distance and cosine similarity, are frequently employed to evaluate the similarity between vectors. The Euclidean distance between two embedding vectors, $\mathbf{v(s_1)}$ and $\mathbf{v(s_2)}$, each with n dimensions representing sentence 1 and sentence 2, is defined as follows:

\begin{equation}
\label{eq:euclidean}
\begin{split}
  d(s_1,s_2) &= {\left\lVert \mathbf{v(s_1)} - \mathbf{v(s_2)} \right\rVert _2} \\
  &=\sqrt{\sum_{i=0}^{n - 1}(v^1_i- v^2_i)^2}
    \end{split}
\end{equation}

Cosine similarity between two vectors $\mathbf{v(s_1)}$ and $\mathbf{v(s_2)}$  is also defined as follows.
\begin{equation}
\label{eq:cosine}
  \begin{split}
  sim(s_1,s_2) &= \frac{\mathbf{v(s_1)} \cdot \mathbf{v(s_2)}}{\left\lVert \mathbf{v(s_1)} \right\rVert _2 \left\lVert \mathbf{v(s_2)} \right\rVert _2} \\
  &= \frac{\sum_{i=0}^{n - 1} (v^1_i v^2_i)}{\sqrt{\sum{{v^1_i}^2}} \sqrt{\sum{{v^2_i}^2}}}
  \end{split}
\end{equation}

In RAG systems, KNN serves as the core retrieval mechanism, where embeddings of queries are matched against a database of document embeddings to retrieve the most relevant context passages; in this context, the top 1 passage was retrieved. The quality and efficiency of KNN retrieval are heavily dependent on the structure of the embedding space, shaped by the underlying model and dimensionalities \cite{b4}. However, KNN suffers from scalability issues, as the time complexity grows linearly with the size of the dataset. Despite these limitations, KNN remains a fundamental technique for embedding-based information retrieval due to its effectiveness and conceptual simplicity.

To address the scalability issue of KNN, Hierarchical Navigable Small World
Algorithm (HNSW) \cite{b8} is commonly employed to accelerate retrieval while maintaining high performance.  The HNSW algorithm is fast and efficient while being able to maintain a high accuracy. It utilizes Skip List and Navigable Small World techniques to create a hierarchical graph-based structure that offers logarithmic time complexity in many cases. This method improves existing graph-based techniques by focusing on the challenges of scalability, accuracy, and efficiency in high-dimensional space. In this work, though not the main method, HNSW served as an experimental method for comparison purposes as well.


After conducting preliminary experiments and comparisons, we chose truncation-based dimensionality reduction over PCA as our method of choice, due to its lower computational overhead despite achieving similar accuracy. In addition to dimensionality reduction, we also explored various state-of-the-art searching mechanisms, including ANNOY \cite{b12, b13}, IVF-PQ \cite{b14, b15}, and HNSW, in the early stages of this work. While ANNOY and IVF-PQ demonstrated fast performance, they fell short in terms of accuracy compared to KNN. HNSW, on the other hand, achieved high accuracy in multiple scenarios, but its graph structure construction required excessive time and complexity, making a fair comparison of time efficiency challenging. As a result, we adopted traditional KNN with truncation as our baseline, allowing us to effectively evaluate the efficacy of our proposed progressive truncation approach and provide a more accurate assessment of its performance.

\section{Methodology}

\subsection{Dataset}

In order to observe the performance of a RAG system in a real world scenario, 1 M English corpus was utilized from the dbpedia-openai-1M-1536-angular dataset \cite{b9}. For the ground truth, 2,500 sentences were randomly selected from the 1 million documents. However, a few problematic sentences that don’t contain actual subjects or content were observed and needed to be filtered out. After the data processing, 2,470 cleaned ground truths were collected. Each document was then passed as context to OpenAI ChatGPT-4o API \cite{b10} to generate a corresponding question. In the end, 2,470 pairs of query and context were prepared for embedding and evaluation.

\subsection{Embedding}

The 1 million documents and the 2,470 queries were embedded using the two embedding models as described in Table \ref{tab_emb}. 

\begin{table}[ht]
\caption{Embedding Models}
\label{tab_emb}
\begin{center} 
\scalebox{0.95}
{
\begin{tabular}{|c|c|c|}
\hline
\rule[-1ex]{0pt}{3.5ex}  Model & Dimension \\\hline\hline
\rule[-1ex]{0pt}{3.5ex}  Alibaba-NLP gte-Qwen2-7B-instruct & 3584 \\\hline
\rule[-1ex]{0pt}{3.5ex}  OpenAI text-embedding-3-large & 3072 \\\hline
\end{tabular}\vspace{-20pt}
}
\end{center}
\end{table}

\subsection{Truncated Retrieval}
We investigated how retrieval performance varies across a wide array of embedding dimensions, ranging from lower dimensions such as 64 all the way to the embedding models’ full output sizes of 3072 and 3584 for OpenAI and Alibaba-NLP respectively. 
We reduced the dimensions of the embedding vectors to 2$^n$ where 4 $\leq$ n $\leq$ 11  (Dim = 16, 32, 64, … 2048) through truncation and PCA separately for each embedding model. 
We used Exact Nearest Neighbors search (1-NN) and HNSW to build indices from the vector database based on the reduced dimension. Then, we retrieved the index of top-1 match for each query based on Euclidean distance.
We extracted the exact document of the retrieved indices from the vector database and compared them to the ground truth. The entire process is shown in Fig.\ref{fig:retrieval}.
It was determined from our findings that we should avoid working further with HNSW and PCA for this research. While accuracy levels were still competitive for HNSW in comparison to other search algorithms, there was a major drawback in the time to build, which is a valuable aspect of retrieval. Additionally, our decision to move away from PCA was due to the lack of significant difference between PCA and Truncation. In fact, truncation dimensionality reduction proved to be slightly more favorable to a higher retrieval accuracy than PCA.

\subsection{Progressive Retrieval}

\begin{figure}[ht]
\centerline{\includegraphics[width=0.95\linewidth]{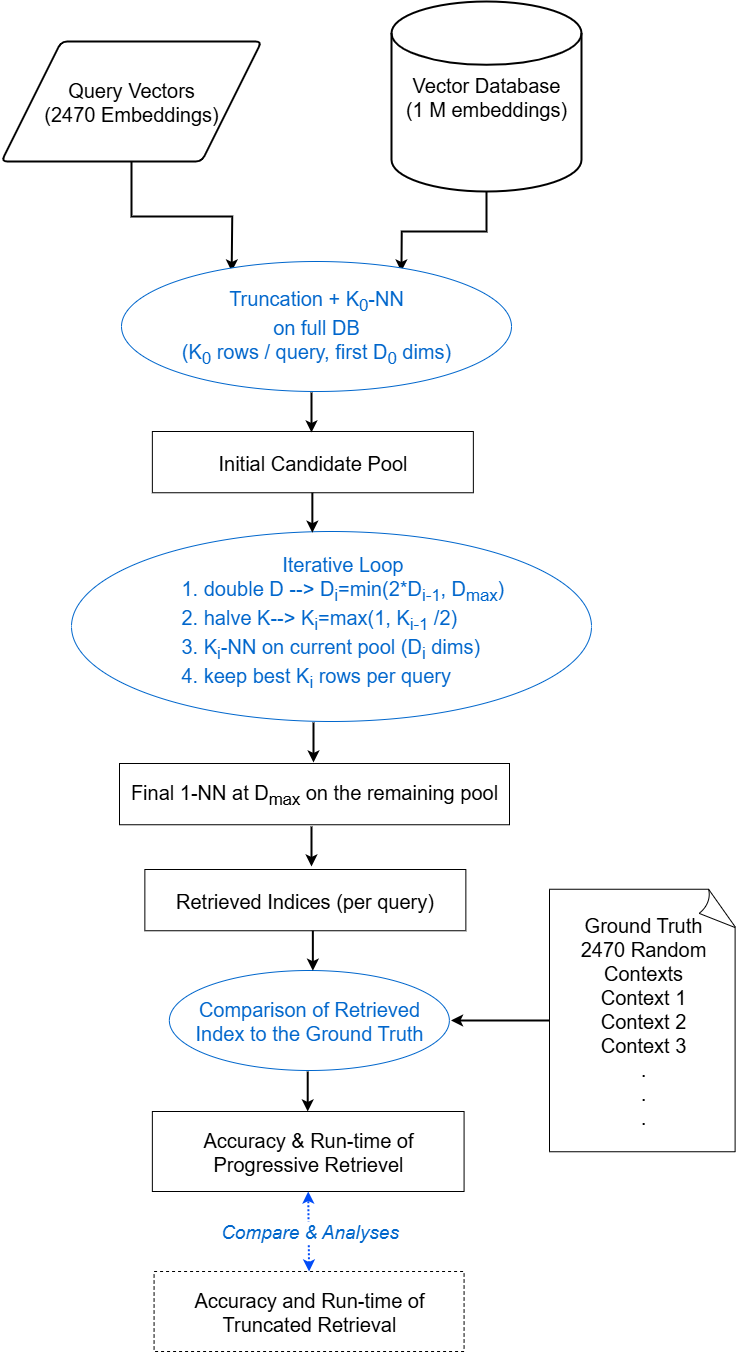}}
\caption{Progressive Retrieval workflow}
\label{fig:progressive}
\end{figure}

We propose a progressive retrieval method that filters candidate documents with multi-stage dimensions. This method aims to reduce the overall query time while maintaining high-levels of retrieval accuracy. By comparing this method against conventional single-stage, we demonstrate that it is possible to achieve efficient and accurate semantic search at a large scale, making it a practical solution for modern RAG pipelines.

The new method requires a few parameters such as initial K, starting dimension, and max dimension to begin. More specifically, initial K is the number of the nearest neighbors that are retrieved for each query in the first search, and they are collected and saved in the same candidate pool, so the duplicate neighbors will be removed. Starting dimension is the number of columns of the database and query vectors where searching with initial K will be used for the entire dataset. Max dimension is the last stage of searching where it performs 1-NN search on the remaining data pool.

Once the parameters are determined, the search would perform on the starting dimension and the entire vector database. Then, it enters a loop, and at each iteration we double the starting dimension and check whether the new dimension coverage is less than the declared max dimension. If the doubled dimension coverage is less than the max dimension, it halves the last K per query (minimum 1), and search with the updated K value on the new dimension coverage and surviving rows. This process continues until the doubled dimension is greater than or equal to the provided max dimension. 
Once the doubled dimension reaches or exceeds the target max dimension, the loop stops and a final 1-NN is run on the remaining candidates and the max dimension.
Fig.\ref{fig:progressive} shows the entire process of the Progressive Retrieval.

\subsection{Metrics}
Our work was conducted in  Google Colab, which means that there could be potential latency during the time measurements due to the network issue. Taking this into account, each experiment was conducted 10 times, and for each iteration, the time was recorded and observed.
\begin{itemize}
\item Top-1 Accuracy: To find the closest neighbor for each query, Scikit-learn’s Nearest Neighbors \cite{b11} (algorithm="brute", metric="euclidean") function is used to calculate the distances. The retrieved indices are then used to extract the actual document from the original 1 million rows, and those documents are then compared to the ground truth to generate the Top-1 accuracy.
\item Median search time: To avoid any outlier in the time measurements and accurately describe the search time, the median result is calculated and used as the Run-Time for each experiment.
\end{itemize}

The accuracy and time consumptions of Truncated Retrieval are used as the baseline, and they will be used to compare the proposed solution - Progressive Retrieval. Both results will be plotted to observe the speed and accuracy of the new method.

\begin{table}[ht]
\caption{Accuracy and Runtime for Truncated Retrieval for the embedding by gte-Qwen2-7B-instruct}
\label{tab_truncated1}
\begin{center} 
\scalebox{0.9}
{
\begin{tabular}{|c|c|c|c|}
\hline
\rule[-1ex]{0pt}{3.5ex}  Dimension & Accuracy (\%) & Run-Time (sec) \\\hline\hline
\rule[-1ex]{0pt}{3.5ex}  16 & 6.56 & 2.70 \\\hline
\rule[-1ex]{0pt}{3.5ex}  32 & 39.55 & 3.01 \\\hline
\rule[-1ex]{0pt}{3.5ex}  64 & 78.42 & 3.83 \\\hline
\rule[-1ex]{0pt}{3.5ex}  128 & 88.79 & 5.71 \\\hline
\rule[-1ex]{0pt}{3.5ex}  256 & 92.79 & 9.38 \\\hline
\rule[-1ex]{0pt}{3.5ex}  512 & 93.81 & 16.44 \\\hline
\rule[-1ex]{0pt}{3.5ex}  1024 & 94.49 & 29.70 \\\hline
\rule[-1ex]{0pt}{3.5ex}  2048 & 94.82 & 57.49 \\\hline
\rule[-1ex]{0pt}{3.5ex}  3072 & 94.98 & 87.50 \\\hline
\rule[-1ex]{0pt}{3.5ex}  3584(Full) & 95.02 & 99.36 \\\hline
\end{tabular}\vspace{-20pt}
}
\end{center}
\end{table}

\section{Experimental Results}

\subsection{gte-Qwen2-7B-instruct}

Table \ref{tab_truncated1} shows the accuracy and Run-Time for the corresponding dimensions for Truncated Retrieval using the embedding of gte-Qwen2-7B-instruct from Alibaba-NLP. We experimented from dimension 16 to 512 with an interval of 8 in addition to 1024, 2048 and 3584 (max dimension). However, only 2$^n$ dimensions are shown in Table \ref{tab_truncated1} to avoid information overloading. The Run-Time for the Truncated Retrieval is linear, which is expected.

\begin{table}[]
\caption{Comparison of Truncated and Progressive Retrieval for gte-Qwen2-7B-instruct embedding} 
\label{tab_comp1}
\begin{tabular}{|c|c|c|c|c|c|}
\hline
\begin{tabular}[c]{@{}l@{}}Trunc. \\Dim \end{tabular}& \begin{tabular}[c]{@{}l@{}}Trunc. \\ Acc\end{tabular} & \begin{tabular}[c]{@{}l@{}}Trunc. \\runtime\end{tabular}   & \begin{tabular}[c]{@{}l@{}}Prog. config\\(Ds, Dm, K)\end{tabular} & \begin{tabular}[c]{@{}l@{}}Prog. \\Acc \end{tabular}& \begin{tabular}[c]{@{}l@{}} Prog. \\runtime\end{tabular} \\\hline
\rule[-1ex]{0pt}{3.5ex}  256 & 92.79 & 9.38 & (128, 512, 128) & 92.79 & 8.94\\\hline
\rule[-1ex]{0pt}{3.5ex}  512 & 93.81 & 16.44 & (128, 2048, 16) & 93.89 & 7.39\\\hline
\rule[-1ex]{0pt}{3.5ex}  1024 & 94.49 & 29.70 & (128, 3584, 64) & 94.49 & 12.18\\\hline
\rule[-1ex]{0pt}{3.5ex}  2048 & 94.82 & 57.49 & (256, 3584, 16) & 94.82 & 12.17\\\hline
\rule[-1ex]{0pt}{3.5ex}  3584 & 95.02 & 99.36 & (512, 3584, 16) & 95.02 & 20.63\\\hline
\end{tabular}
\end{table}

The progressive Retrieval experiment includes a starting dimension of 64, 128, 256, 512, 1024, 2048, initial K of 4, 8, 16, 32, 64, 128, 256, 512, 1024, and max dimension of 128, 256, 512, 1024, 2048, 3584.
Overall Run-Time is faster than that of Truncated Retrieval while maintaining good accuracy. To better compare the accuracy and Run-time, we selected 5 cases from Truncated Retrieval and also chose fastest configurations (starting dimension, max dimension and starting K) of Progressive Retrieval, which has the same or close-to-same accuracy. The results are gathered and presented in Table \ref{tab_comp1} as shown. Progressive Retrieval is around two times faster than the Truncated Retrieval from accuracy 93.8 to 94.49. Moreover, the proposed method is almost 5x faster at Dimension 2048 and 3584 from accuracy 94.82 to 95.02.

\begin{figure}[ht]
\centerline{\includegraphics[width=0.9\linewidth]{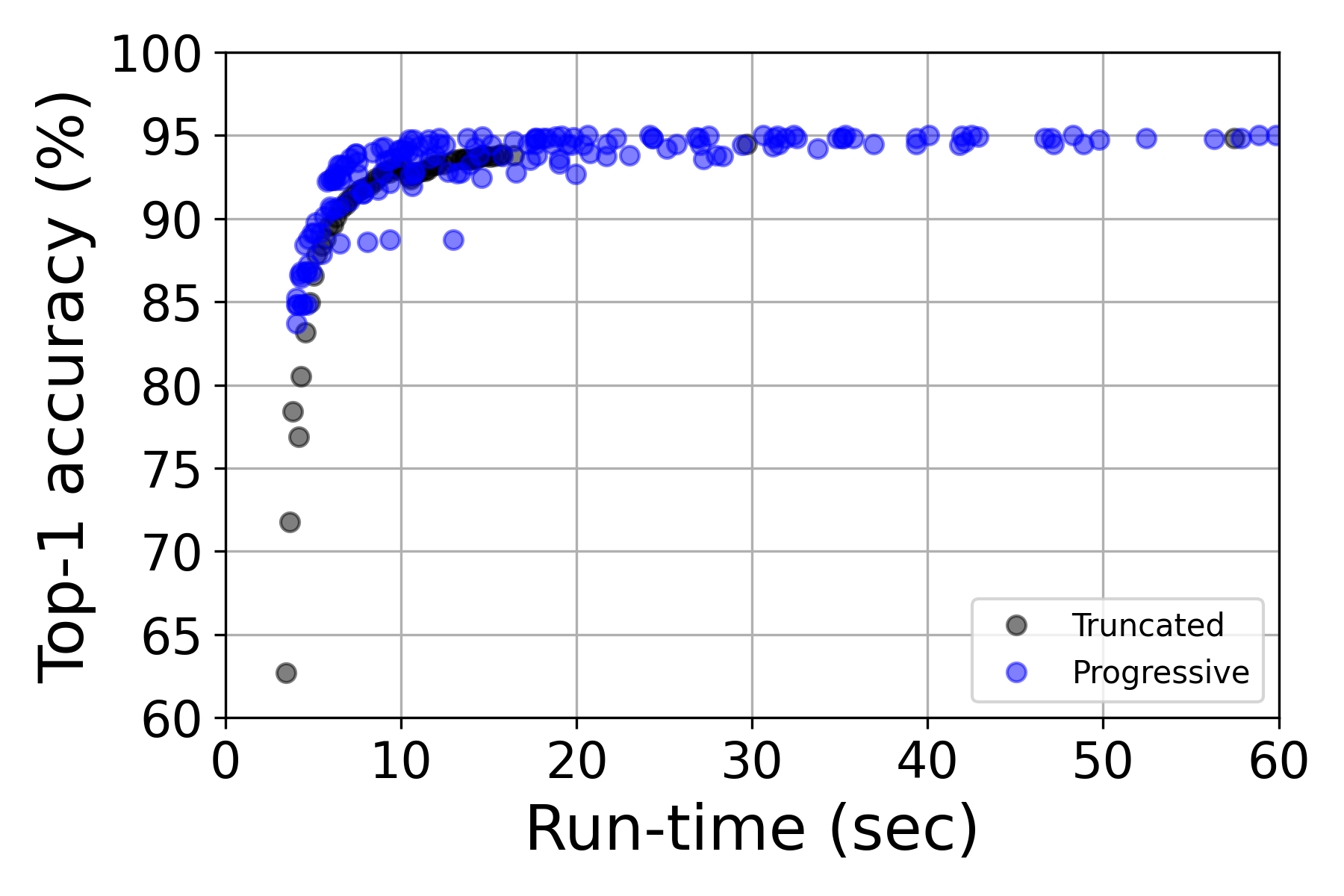}}
\caption{Accuracy vs. Runtime of Retrieval on embedding vectors by gte-Qwen2-7B-instruct }
\label{fig:acc_time1}
\end{figure}

Next, we plot the entire experimental results of Truncated and Progressive Retrieval where the X-axis is Run-time (sec) and Y-axis is Top-1 accuracy (\%). 
Fig.\ref{fig:acc_time1} shows that the majority of the Progressive Retrieval (blue points) are above the Truncated Retrieval (gray points) which suggests most results reach higher accuracy with less run-time.

\subsection{text-embedding-3-large}

Table \ref{tab_truncated2} shows the accuracy and Run-Time for the corresponding dimensions for Truncated Retrieval using the embedding of text-embedding-3-large from OpenAI. We experimented from 16 to 512 dimension by an increment of 8 dimensions. After 512 dimensions, it follows the 2$^n$ dimension pattern until it reaches the maximum dimension, which is 3072. There are 65 cases for Truncated Retrieval. Table \ref{tab_truncated2} is the simple version of the result, which contains only 2$^n$ dimensions and the maximum dimension.

\begin{table}[ht]
\caption{Accuracy and Runtime for Truncated Retrieval for the embedding by text-embedding-3-large}
\label{tab_truncated2}
\begin{center} 
\scalebox{0.9}
{
\begin{tabular}{|c|c|c|c|}
\hline
\rule[-1ex]{0pt}{3.5ex}  Dimension & Accuracy (\%) & Run-Time (sec) \\\hline\hline
\rule[-1ex]{0pt}{3.5ex}  16 & 3.32 & 2.82 \\\hline
\rule[-1ex]{0pt}{3.5ex}  32 & 29.35 & 3.29 \\\hline
\rule[-1ex]{0pt}{3.5ex}  64 & 70.73 & 4.16 \\\hline
\rule[-1ex]{0pt}{3.5ex}  128 & 88.18 & 6.19 \\\hline
\rule[-1ex]{0pt}{3.5ex}  256 & 92.02 & 9.68 \\\hline
\rule[-1ex]{0pt}{3.5ex}  512 & 93.40 & 16.82 \\\hline
\rule[-1ex]{0pt}{3.5ex}  1024 & 93.85 & 30.85 \\\hline
\rule[-1ex]{0pt}{3.5ex}  2048 & 94.17 & 59.86 \\\hline
\rule[-1ex]{0pt}{3.5ex}  3072 & 94.45 & 80.31 \\\hline
\end{tabular}\vspace{-20pt}
}
\end{center}
\end{table}

The Progressive Retrieval conducted in a range of dimensions starting from 128 to max dimension 3072, and the start k section contains 16, 32, 64, and 128. 
Table \ref{tab_comp2} is the comparison between baseline Truncated and Progressive Retrieval based on accuracy. The example this table uses for Progressive Retrieval is the best case, which means under the same or relatively similar accuracy, the best run time will be selected.


An interesting trend emerges when analyzing the results from Table \ref{tab_comp1} and Table \ref{tab_comp2}. The run-time and accuracy of Truncated Retrieval exhibit a linear trend, which is expected given that the naive search is performed on the entire one-million documents at each dimension. As the dimensions increase, the query time predictably rises. In contrast, the run-time of Progressive Retrieval deviates from this linear pattern. This is because the initial step, which involves querying the one-million document embeddings at a low dimension, dominates the query time. Although the information is not included in the tables due to space constraints in this paper, when the starting dimension is 128, the query time remains relatively consistent at around 10 seconds. The starting dimension, along with the initial K value, plays a significant role in determining the total query time. Notably, the subsequent steps of Progressive Retrieval are computationally less expensive, yet they substantially improve accuracy. Overall, the query time for Progressive Retrieval is generally shorter than that of Truncated Retrieval, particularly when compared to Truncated Retrieval using high-dimensional vectors.

\begin{table}[]
\caption{Comparison of Truncated and Progressive Retrieval for text-embedding-3-large embedding} 
\label{tab_comp2}
\begin{tabular}{|c|c|c|c|c|c|}
\hline
\begin{tabular}[c]{@{}l@{}}Trunc. \\Dim \end{tabular}& \begin{tabular}[c]{@{}l@{}}Trunc. \\ Acc\end{tabular} & \begin{tabular}[c]{@{}l@{}}Trunc. \\runtime\end{tabular}   & \begin{tabular}[c]{@{}l@{}}Prog. config\\(Ds, Dm, K)\end{tabular} & \begin{tabular}[c]{@{}l@{}}Prog. \\Acc \end{tabular}& \begin{tabular}[c]{@{}l@{}} Prog. \\runtime\end{tabular} \\\hline
\rule[-1ex]{0pt}{3.5ex}  256 & 92.02 & 9.68 & (128, 256, 128) & 92.02 & 10.50\\\hline
\rule[-1ex]{0pt}{3.5ex}  512 & 93.40 & 16.82 & (256, 512, 16) & 93.40 & 10.74\\\hline
\rule[-1ex]{0pt}{3.5ex}  1024 & 93.85 & 30.85 & (128, 2048, 32) & 93.85 & 9.99\\\hline
\rule[-1ex]{0pt}{3.5ex}  2048 & 94.17 & 59.86 & (128, 3072, 64) & 94.17 & 14.04\\\hline
\rule[-1ex]{0pt}{3.5ex}  3072 & 94.45 & 80.31 & (256, 3072, 64) & 94.45 & 20.50\\\hline
\end{tabular}
\end{table}

\begin{figure}[ht]
\centerline{\includegraphics[width=0.9\linewidth]{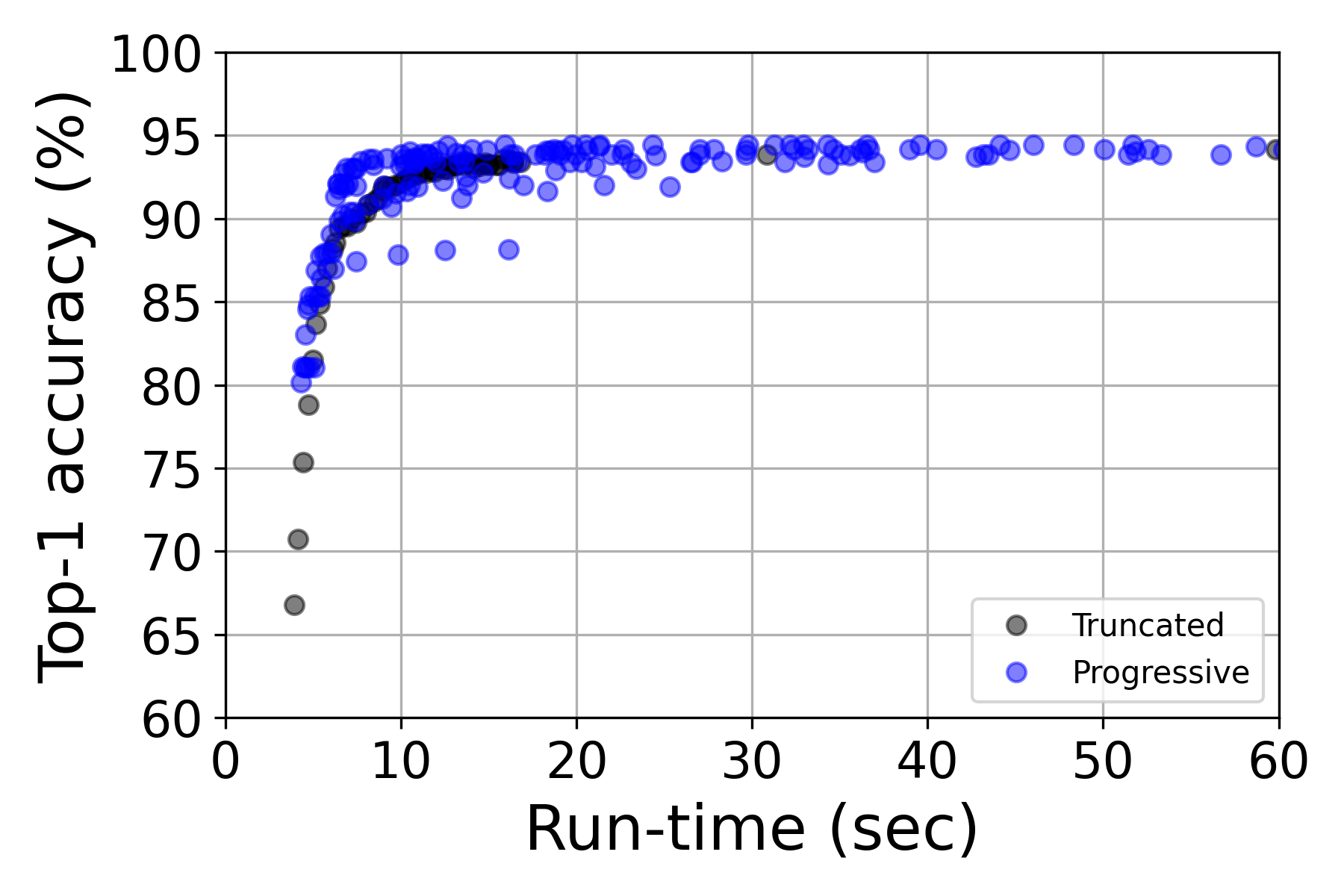}}
\caption{Accuracy vs. Run-time of Retrieval on embedding vectors by text-embedding-3-large }
\label{fig:acc_time2}
\end{figure}

After all data is found, all points are plotted into a graph. Fig.\ref{fig:acc_time2} is the complete graph that contains all of the data points for both methods. 
It should be noted that we excluded some of our results for when the dimensions fell below 64 in the case of gte-Qwen2-7B-instruct due to their consistently poor performance for both the Truncated Retrieval as well as for Progressive Retrieval. This reinforces the notion that there is a lower bound of dimensionality beneath which both Truncated and Progressive Retrieval methods suffer in retrieval quality, and such dimensions should be avoided unless speed is the only concern.

\section{Analysis}

Our results demonstrate that the Progressive Retrieval method often reduces query time substantially while preserving high accuracy compared with the Truncated Retrieval Method.
This can be seen in both embedding models and across many of the tested dimensions. In the case of gte-Qwen2-7B-instruct embeddings, the naive KNN achieved its peak accuracy of 95.02\% at the full dimensionality of 3584, which was to be expected. 
However, the run-time was extremely long at nearly 100 seconds. Progressive Retrieval method, on the other hand, was able to achieve the same accuracy in only 20.63 seconds, indicating a nearly 5x speedup with no loss in performance in Table \ref{tab_truncated1}.

In general, most of our Progressive Retrieval method outperformed or matched the Truncated Retrieval method across the accuracy-runtime tradeoff. As can be seen from Fig.\ref{fig:acc_time1}, a large number of the Progressive Retrieval results lie above the Truncated Retrieval curve, which suggests that there is better accuracy for the same time in many cases. What is important to note is that while not every configuration of the Progressive Retrieval yielded results above the Truncated Retrieval curve, there is clear evidence that with Progressive Retrieval the time it takes to reach certain accuracies is drastically reduced. 
Moreover, the results indicate that Progressive Retrieval offers a more efficient approach in scenarios where a slightly lower, yet still competitive, accuracy is acceptable, as it can achieve this in a significantly shorter timeframe.

A notable trend emerged from our work with the Progressive Retrieval, which was that the starting dimension parameter had the greatest impact on both runtime and accuracy, followed by the starting K value, and lastly by the maximum dimension. 
We attribute this trend to be a cause of the nature of how Progressive Retrieval is performed. The largest portion of time and effort will be performed on the starting dimension. This is because the starting dimension search for relevant documents is done on the entire 1 million rows of data. The proceeding steps in the Progressive Retrieval work off of a candidate pool that is much smaller each iteration, meaning that, while the dimensions are increasing, the number of documents that are being queried is drastically reduced. Extending on this idea, as we reach the maximum dimension to perform the top K=1 retrieval, there are a very small number of candidate documents, making the retrieval time for that particular step very fast. As noted previously, the K value can play a significant role in accuracy and runtime as well, but the main influence on how long the query will take and the accuracy achieved is contributed to the initial document query on the entire dataset. Subsequent steps in the Progressive Retrieval method take less time.


Our results also reveal an important characteristic of Progressive Retrieval: its accuracy consistently falls within the range defined by the minimum and maximum dimensions set. This means that Progressive Retrieval always outperforms the lower dimensional bound in terms of accuracy, but never surpasses the accuracy of the upper dimension. Notably, despite this bounded accuracy, Progressive Retrieval achieves significantly faster retrieval times compared to the baseline, which is a crucial advantage in document retrieval applications, particularly in RAG systems.



\section{Conclusion}

In this study, we explored the impact of embedding dimensionality on document retrieval performance in RAG systems, focusing on both accuracy and computational efficiency at scale. Using a dataset of one million documents and two state-of-the-art embedding models from OpenAI and Alibaba-NLP, we systematically evaluated retrieval accuracy across a wide spectrum of vector dimensionalities. Our work involved 2470 query-document pairs, allowing us to measure top-1 retrieval precision under controlled experimental conditions using KNN as the retrieval mechanisms.

To address the growing computational cost of high-dimensional similarity search, we introduced a novel Progressive Retrieval method. This method incrementally refines the candidate pool using a hierarchy of KNN searches, starting from reduced-dimensional representations and progressing to the full target dimensionality. By filtering candidates in early stages and only applying larger dimensionality searches to a smaller subset of our dataset, the Progressive search method offers a scalable and efficient alternative to brute-force retrieval in high-dimensional vector spaces.

Our findings demonstrate that embedding dimensionality plays a critical role in balancing retrieval precision and the speed at which the retrieval was performed. In particular, we show that while aggressive dimensionality reduction can be detrimental to accuracy, our progressive method enables a middle ground that maintains high precision with reduced query times. This has important implications for the design of large-scale RAG systems, where trade-offs between model complexity and real-time retrieval performance must be carefully managed.

As embedding models continue to grow in complexity and document corpora expand in size, scalable retrieval strategies like our Progressive Retrieval method will be essential. Future work will explore more advanced candidate filtering techniques, integration with Approximate Nearest Neighbor Search methods \cite{b13, b16}, and potential application beyond document retrieval. We hope that with the  findings of this study, further exploration of optimizing RAG systems can be done and continue to grow in accordance with the increasing complexity of LLMs in the field of NLP today.

\vspace{12pt}
\color{red}

\end{document}